\documentclass[fleqn,10pt]{wlscirep}
\usepackage[utf8]{inputenc}
\usepackage[T1]{fontenc}
\usepackage{lineno}

\title{The Goldilocks Principle of Learning Unitaries by Interlacing Fixed Operators with Programmable Phase Shifters on a Photonic Chip}

\author[1,+]{Kevin Zelaya}
\author[1,2,+]{Matthew Markowitz}
\author[1,2,+,*]{Mohammad-Ali Miri}
\affil[1]{Department of Physics, Queens College of the City University of New York, Queens, New York 11367, USA}
\affil[2]{Physics Program, The Graduate Center, City University of New York, New York, New York 10016, USA}

\affil[*]{mmiri@qc.cuny.edu}

\affil[+]{these authors contributed equally to this work}


\begin{abstract}
Programmable photonic integrated circuits represent an emerging technology that amalgamates photonics and electronics, paving the way for light-based information processing at high speeds and low power consumption. Programmable photonics provides a flexible platform that can be reconfigured to perform multiple tasks, thereby holding great promise for revolutionizing future optical networks and quantum computing systems. Over the past decade, there has been constant progress in developing several different architectures for realizing programmable photonic circuits that allow for realizing arbitrary discrete unitary operations with light. 
Here, we systematically investigate a general family of photonic circuits for realizing arbitrary unitaries based on a simple architecture that interlaces a fixed intervening layer with programmable phase shifter layers. 
We introduce a criterion for the intervening operator that guarantees the universality of this architecture for representing arbitrary $N \times N$ unitary operators with $N+1$ phase layers. We explore this criterion for different photonic components, including photonic waveguide lattices and meshes of directional couplers, which allows the identification of several families of photonic components that can serve as the intervening layers in the interlacing architecture. Our findings pave the way for efficiently designing and realizing novel families of programmable photonic integrated circuits for multipurpose analog information processing.
\end{abstract}

\begin{document}

\flushbottom
\maketitle
%
%
\thispagestyle{empty}

Levering the unique properties of light to perform computations in novel ways is a subject with a long history \cite{joannopoulos1997photonic,yan2009nanowire}. Although an all-optical processor for universal computing seems to be a far reach goal, photonics can provide exciting opportunities for unconventional computing building on analog logic and with novel information processing configurations. What makes photonics an intriguing option for unconventional computing are exotic potentials such as the intrinsically high speed and low energy consumption, the capabilities for massive parallelization, and long-range interactions. Nevertheless, a significant challenge in optical computing lies in the absence of appropriate computing paradigms, methods, and algorithms that harness the unique capabilities of this technology to develop efficient and application-specific photonic processors. In particular, matrix-by-vector multiplication is one of the most basic mathematical operations that lies at the core of various tasks ranging from optical convolution schemes~\cite{shen2017deep,zelaya2023integrated} and matrix eigenvalue solver~\cite{liao2022matrix} to novel optical memristors~\cite{mao2019photonic,youngblood2023integrated} and optical artificial neural networks \cite{ashtiani2022chip,liao2023integrated}. In the past decade, with rapid technological progress, there has been a resurrection in efforts devoted to developing such programmable photonic integrated circuit that performs matrix-vector multiplication~\cite{Saygin2020,Zhou22, Pastor21}. The utility of such a device as an energy-efficient photonic accelerator in conjunction with electronic processors appears to be a distant possibility, considering the inherent difficulties associated with scaling and precision. However, there is no doubt that an on-chip programmable photonic matrix-vector multiplier can create exciting opportunities in classical and quantum computing through various applications that range from quantum information and quantum transport simulations ~\cite{madsen2022quantum,Harris17,slussarenko2019photonic}, to optical signal processing \cite{notaros2017programmable}, neuromorphic computing~\cite{Xu23neuromorphic}, and optical neural networks~\cite{shen2017deep,zhu2022space,li2024high}, as well as putting forward a platform for rapid prototyping of linear multiport photonic devices \cite{Tang:22,xu2022parallel}.

Indeed, the optical realization of arbitrary unitary operations has been known since the seminal paper of Reck \textit{et al.}~\cite{Reck94}, which originally concerned free space optics but successfully translated to photonic integrated circuits by Miller~\cite{Miller2012, Miller2013a, Miller2013b}. This architecture builds on breaking down unitary matrices of any order into lower-dimension unitary matrices, which ensures the existence of an optical realization through two fundamental building blocks that are beam splitters (couplers) and phase shifters. Despite its generality, this method uses a pyramid-shaped array of Mach-Zehnder interferometers (MZI), which is impractical for larger implementations because the number of beam splitters grows quadratically with the number of ports. In turn, Clements \textit{et al.}~\cite{Clements16} introduced an alternative and symmetric rectangular-shaped array, resulting in a device with half the total optical depth and, consequently, more loss-tolerant. Such a rectangular array has been proved robust enough to create photonic realizations of Haar-random matrices~\cite{Burgwal2017}. Further unitary realizations related to other mesh geometries have been explored in~\cite{Shokraneh2020, Mojaver2023}, as well as topological photonic lattices with hexagonal-shaped arrays of MZI~\cite{On2023, Wang2019}. In turn, free-space propagation setups have been devised based on plane-light conversion~\cite{Labroille14, Morizur10} and using diffractive surface layers~\cite{Kulce2021}. 

While the latter devices originally consisted of bulky optical components, the principle has recently been applied to on-chip structures as well. Recent studies have explored the use of particular transfer matrices (henceforth called $F$) alternating with phase mask layers to obtain an arbitrary unitary transformation~\cite{Tanomura20, Saygin2020, Tanomura22a, Pastor21, markowitz2023universal}. Pastor et al.~\cite{Pastor21} considers wave propagation in multimode slab waveguides to implement a Discrete Fourier transform (DFT) as their transformation $F$. They showed that an arbitrary transformation could be performed when $6N+1$ phase layers and $6N$ DFT elements, where $N$ is the number of ports. Tanomura \textit{et al.}~\cite{Tanomura20} interleave the phase masks with multimode interference couplers connected with single-mode waveguides and use simulated annealing optimization to argue for well-approximated conversions when $M \approx N$. Fully functional unitary four-, eight-, ten-, and twelve-port devices have been proposed and manufactured~\cite{ribeiro2016demonstration,taballione20198,Tang2021,taballione2021}. Moreover, an alternative device using polarization and multiple wavelength degrees of freedom has been considered in~\cite{Tanomura23}. Markowitz and Miri~\cite{markowitz2023universal} have explored similar structures and have found rigorous numerical evidence that interleaved phase arrays and discrete fractional Fourier transform (DFrFT) are indeed universal, whereas the use of Haar-random unitary matrices has been proved to lead to the desired universality~\cite{Saygin2020}. A further waveguide array with varying propagation constants with step-like profiles has been reported~\cite{Skryabin2021}. The interlacing architecture appears to exhibit interesting auto-calibrating properties, which makes it resilient to fabrication errors \cite{Markowitz23Auto}. Furthermore, we recently have shown that the intervening structure can go beyond implementing unitaries to directly implementing arbitrary non-unitary operations when the diagonal matrices are relaxed to leave the unitary circle in the complex domain \cite{markowitz2023learning}. In this sense, by utilizing both amplitude and phase modulations, one can realize a fully programmable device for arbitrary matrix operations~\cite{markowitz2023learning}. This shows an important generalization of the previous results that show by itself an advantage of the interlacing architecture over the mesh geometries.

This manuscript discusses a broad class of universal on-chip photonic architectures based on a layered configuration of phase mask layers as programmable units interlaced with a passive random matrix $F$. It is shown that the proposed interlaced architecture is far more flexible by showing that broad families of matrices $F$ can serve as the fixed intervening operator. The phases are steered to reconstruct a unitary target matrix, provided that $F$ has well-posed properties. Numerical evidence based on rigorous optimization algorithms reveals that universality is reached for dense matrices $F$, while a phase transition in the accuracy of reconstructed $N \times N$ targets occurs at $M=N+1$, with $M$ the total number of phase mask layers. Tests using the discrete Fourier transform (DFT) and discrete fractional Fourier transform (DFrFT) confirm the latter claim, and Haar-random matrices also show outstanding convergence. To generalize the domain of the valid intervening operators, a density criterion is derived so that matrices $F$ can be classified according to their elements to ensure universality. To demonstrate this result, photonic lattices with uniform, nonuniform, and disordered coupling coefficients are considered as photonic realizations for the matrix $F$, while using the proposed density criterion, it is shown that universality is reached for specific length intervals. Furthermore, we explore waveguide coupler meshes as an alternative intervening unit $F$ and determine the minimum number of coupler layers required to guarantee the universality of the interlacing architecture. 

\section*{Results}
\subsection*{Architecture and Mathematical Foundation}

\begin{figure*}[!ht]
\centering
\includegraphics[width=0.75\textwidth]{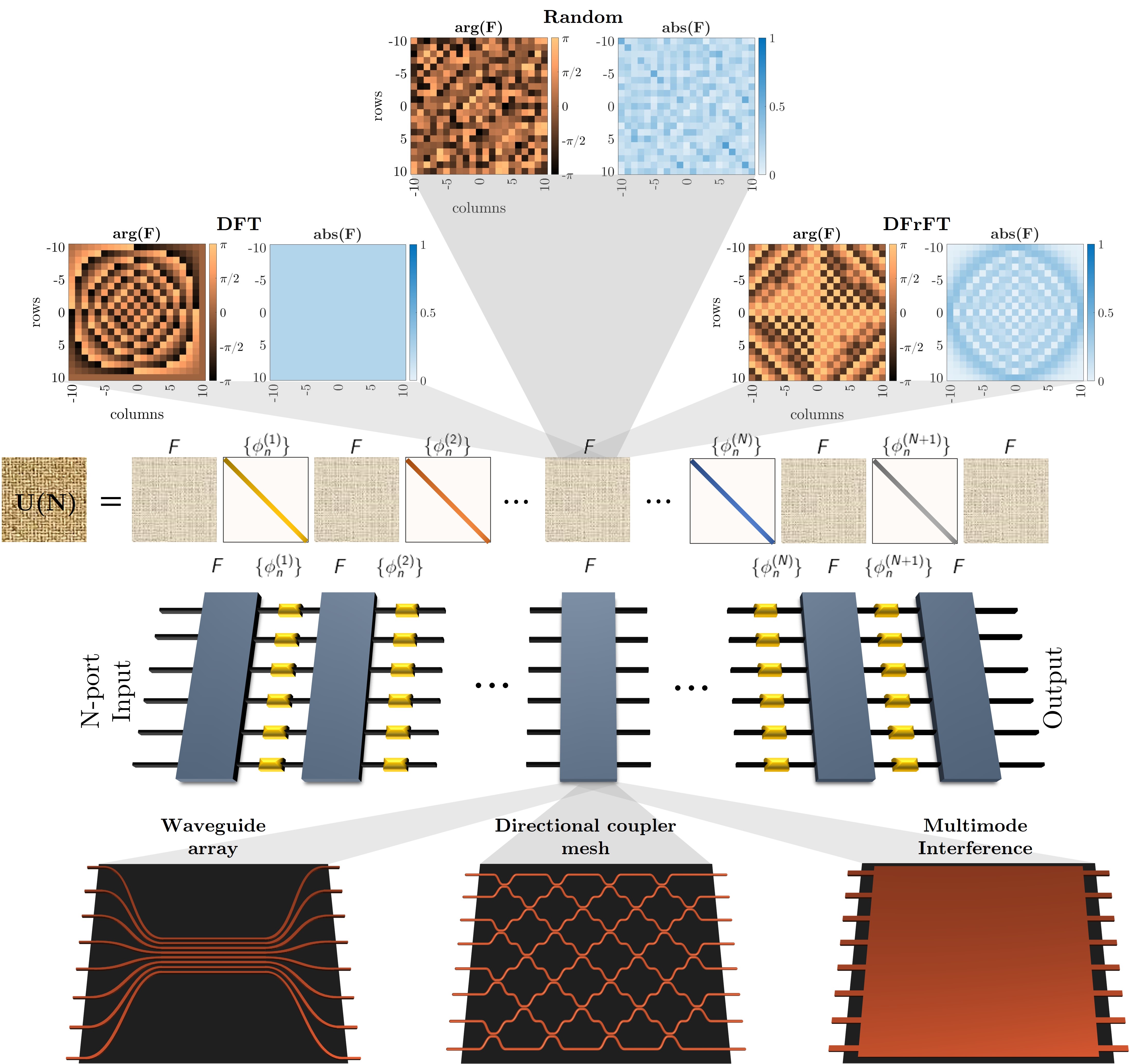}
\caption{\textbf{Universal architecture scheme.} The proposed architecture involving alternating layers of random unitary matrices $F$ and diagonal phase shifts layers (PL) $\{\phi_{n}^{(p)}\}$, with $p=1,\ldots,N+1$. The upper insets depict the modulus and argument of the potential candidates for the unitary matrix $F$, which have been selected as the DFT, DFrFT, and a random unitary matrix. The lower insets illustrate potential photonic implementations to perform the unitary matrix F.}
\label{fig1}
\end{figure*}

Let us consider an arbitrary unitary matrix $\mathcal{U}\in U(N)$, with $U(N)$ the group of unitary matrices $N\times N$. Our goal is to implement a proper factorization of $\mathcal{U}$ in terms of another unitary matrix $F$ to be defined and a set $\{P_{k}\}_{k=1}^{M}$ composed of phase matrices $P_{k}=e^{i D_{k}}$, with $D_{k}=diag(\phi_{1}^{(k)},\ldots,\phi_{N}^{(k)})$ a diagonal matrix with $\phi_{n}^{(k)}\in(0,2\pi]$ for $n\in\{1\ldots,N\}$ and $k\in\{1,\ldots,M\in\mathbb{N}\}$. The factorization proposed here is such that it intercalates $F$ with a phase matrix $P_{k}$ through the relation
\begin{linenomath}
\begin{equation}
\mathcal{U}=FP_{M}F\ldots FP_{1}F.
\label{eq_U}
\end{equation}
\end{linenomath}

This factorization is convenient for optical applications as the phase matrices can be implemented through phase shifter layers, which are the active optical elements of the architecture. In turn, the matrix $F$ (passive optical element) has to be selected so that arbitrary unitary target matrices $\mathcal{U}_{t}$ can be reconstructed with minimal error by adequately tuning the phase shifters $\phi_{n}^{(k)}$. If the latter is achieved, it is said that the \textit{universality property} has been met.
An arbitrary matrix $\mathcal{U}\in U(N)$ requires $N^{2}$ real parameters to be fully defined. Therefore, while performing the factorization, it is vital to consider at least the same number of parameters. For the device proposed in~\eqref{eq_U}, we have $MN$ free parameters in total and expect that $M\geq N$ in order to achieve the desired universality. Although there are some cases where $\mathcal{U}$ has a particular symmetry that reduces the number of parameters, we aim for the general case. The proposed architecture~\eqref{eq_U} and its optical implementation are sketched in Figure~\ref{fig1}. On the one hand, the mathematical structure of the unitary $F$ (see top panels of Figure~\ref{fig1}) can be that of the discrete Fourier transform (DFT), discrete fractional Fourier transform, or simply a Haar-random matrix. On the other hand, the photonic implementation of $F$ can be performed through different approaches (see bottom panels of Figure~\ref{fig1}), such as waveguide arrays~\cite{Wei16,markowitz2023learning}, meshes of directional coupler~\cite{Clements16,Shokraneh2020,Mojaver2023}, and multimode interference (MMI)~\cite{cooney2016analysis,Pastor21}. Here, we focus on architectures based on the first two solutions, the numerical analysis and universality of which are discussed below.
\begin{figure*}
\centering
\includegraphics[width=0.9\textwidth]{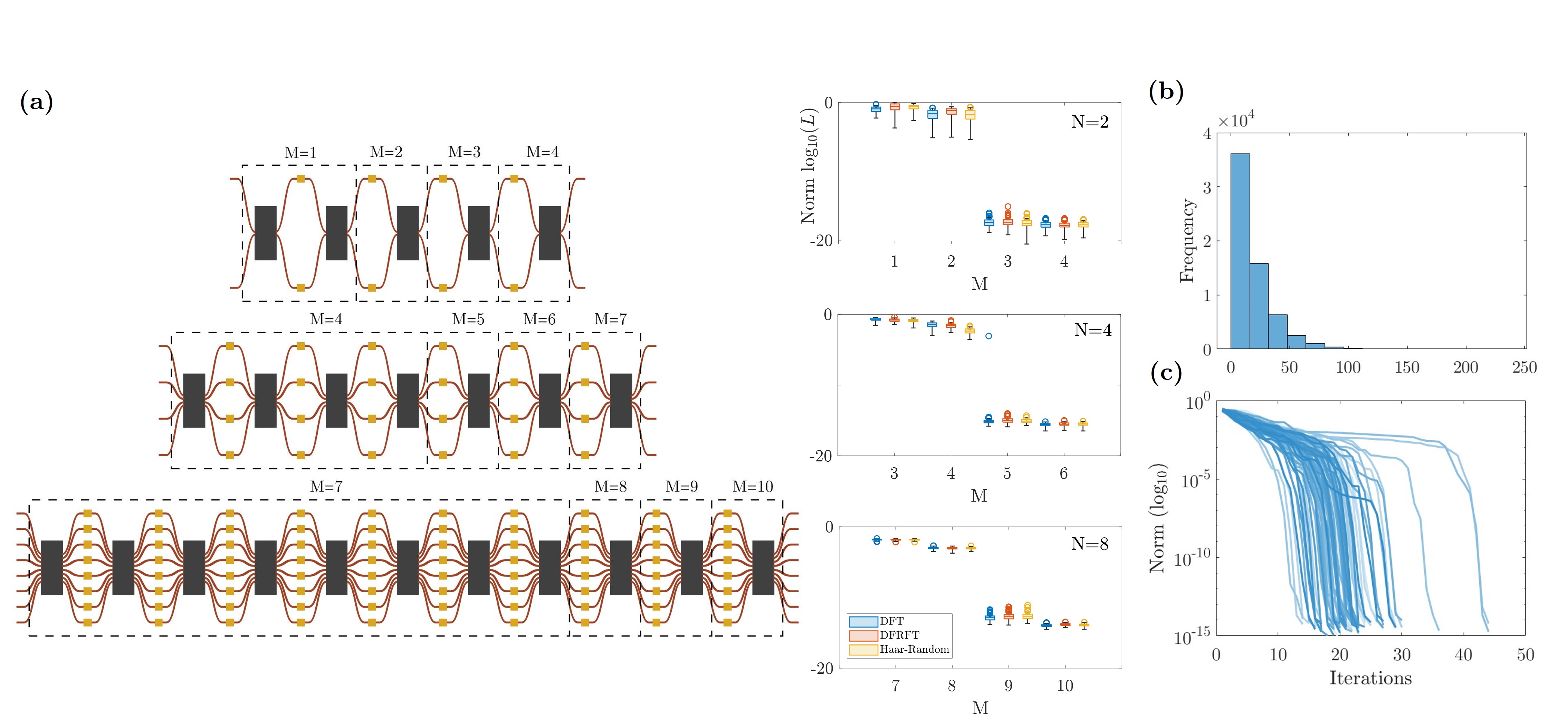}
\caption{\textbf{Numerical universality test.} (a) Architecture depiction (left column) and optimization objective function (right column) for 100 target matrices at various values of M and N. Black boxes denote any possible realization for the $F$ matrix. (b) Multiple trials for $N=8$ and $M=9$ using 250 random $F$ matrices were considered; 250 targets were used for each matrix $F$. Shown is the distribution of the number of LMA runs to achieve a norm lower than the stopping norm of $10^{-10}$, with a maximum of 50 iterations per run. (c) Norm (log${}_{10}$) in terms of the number of iterations for the run with the best norm. Using 100 random matrices $F$, each with a single target matrix.
}
\label{fig2}
\end{figure*}
Layered architectures akin to Eq.~\ref{eq_U} have been numerically validated in previous works using different optical arrays and MZI meshes, where different optimization algorithms such as gradient-descent, stochastic gradient descent, simulated annealing, and basin-hopping have been implemented. See for instance~\cite{Tang2021,Taguchi2023,Skryabin2021,Saygin2020}. In order to demonstrate the universality of this device, we optimize the $NM$ phases for a variety of randomly chosen target matrices $\mathcal{U}_{t}$ generated in accordance with the Haar measure~\cite{mezzadri_how_2007}. The objective function to be minimized, also called \textit{error norm}, is defined by
\begin{linenomath}
\begin{equation}
L = \frac{1}{N^2} ||\mathcal{U}-\mathcal{U}_{t}||^2
\label{Q}
\end{equation}
\end{linenomath}
where $\Vert \cdot\Vert$ stands for the Frobenius definition of the norm, $\mathcal{U}_{t}$ is the target matrix being tested, and $\mathcal{U}$ is the reconstructed matrix using the factorization~\eqref{eq_U}.
The Levenberg-Marquardt (LM) algorithm~\cite{levenberg1944method,Marquardt63} is used to find the minimum of this function. For a given target $\mathcal{U}_{t}$, the phases are randomly initialized between $0$ and $2\pi$. The optimization was performed in MATLAB. In Figure~\ref{fig2}(a), the norm for 100 target unitary matrices is shown for various cases, fixing the default tolerance values to $10^{-10}$. A phase transition occurs between the $M=N$ and $M=N+1$ layers, which is unsurprising given the system becomes over-determined by $N$ parameters. These jumps are larger than reported by Tang et al.~\cite{Tang2017} due to their usage of a probabilistic algorithm (Simulated Annealing) rather than a gradient-based one such as LMA. A downside of gradient-based approaches is they may require many runs with different starting conditions. We can decrease the overhead by using a stopping criteria for the norm along with a maximum iteration for each run of LMA. Using a maximum iteration of 50, we find that we rarely need more than 100 runs in the case $N=8$, $M=9$ to achieve norms less than $10^{-10}$ (Figure ~\ref{fig2}(b)). For systems with a lower number of ports $N$, the distribution skews towards lower values. To more confidently label choices of $F$ which are not Haar-random generated as "bad" mixing layers, we set the maximum number of runs somewhat higher to 250 or 500 as found appropriate. 

\noindent

\subsection*{Density Estimation Criterion}
\label{sec:density}

\begin{figure*}
\centering
\includegraphics[width=0.85\textwidth]{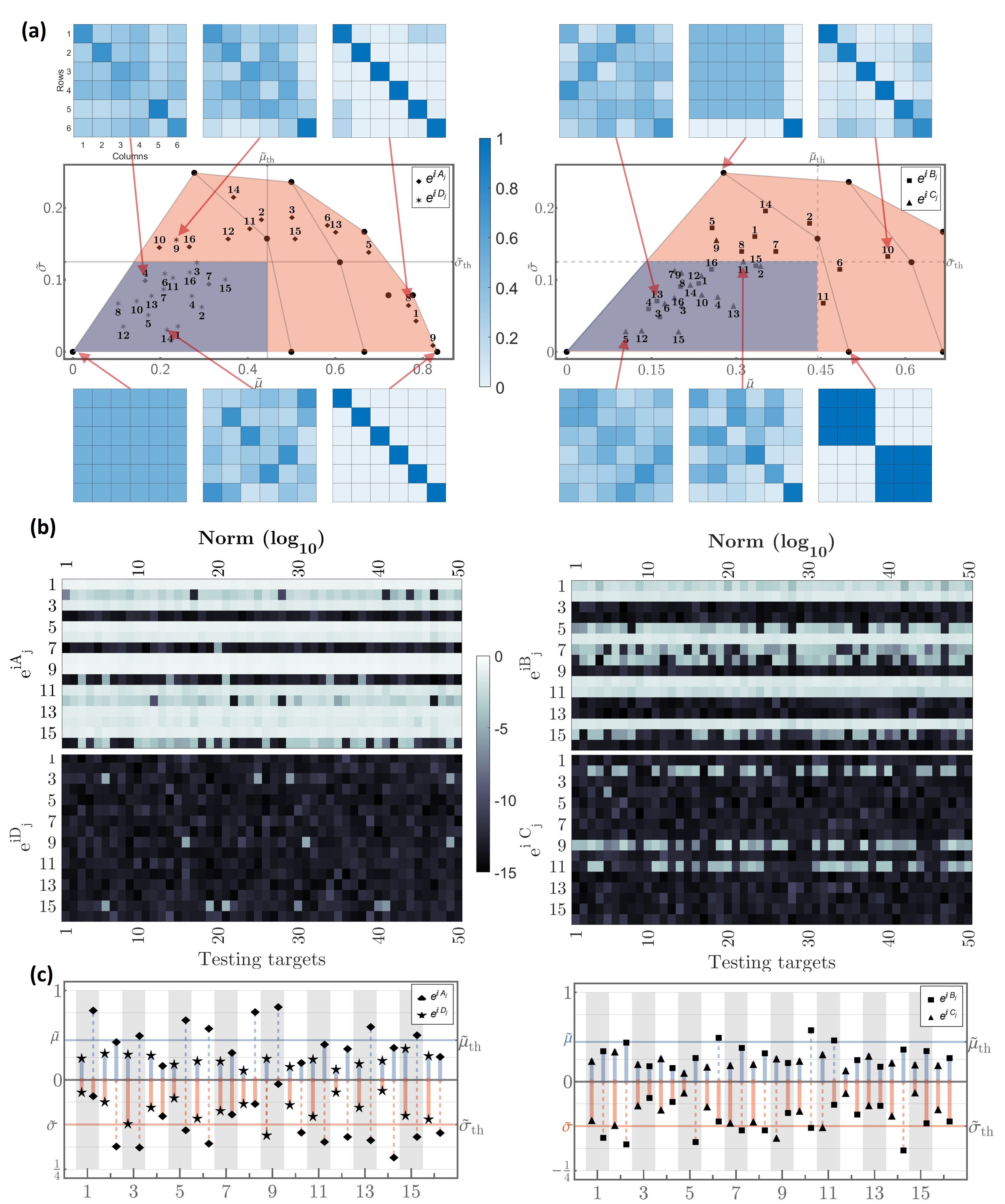}
\caption{\textbf{Density estimation and performance test}. (a) Points $\Vec{R}$ associated with density estimation for the set of unitary matrices $\{e^{A_{j}},e^{D_{j}}\}_{j=1}^{50}$ (left column) and $\{e^{B_{j}},e^{C_{j}}\}_{j=1}^{50}$ (right  column). The blue heat maps denote the absolute value, $\widetilde{\mathcal{U}}$, for some particular choices of unitary matrices. (b) Error norm (log$_{10}$) $L$ in~\eqref{Q} for each unitary matrix under consideration with fifty testing targets per matrix. (c) Mean and standard deviation $N\widetilde{\mu}$ and $N\widetilde{\sigma}$, respectively, related to the density estimation for each unitary matrix in (a). The horizontal blue and red lines denote the universality threshold for $N\widetilde{\mu}$ and $N\widetilde{\sigma}$, respectively.}
\label{fig3}
\end{figure*}

Our preliminary numerical results suggest that dense matrices $F$ lead to a universal architecture, and it is thus desirable to find a proper measure to quantify the density for unitary matrices. Indeed, random matrix theory establishes robust criteria for studying random complex-valued matrices at the limit of large dimensions based on the singular value decomposition (SVD) of the same~\cite{Livan18}. On the one hand, in the current setup, we focus on unitary matrices of relatively small size, as the number of ports in our architecture is not necessarily long enough to take into account the asymptotic analysis of random matrix theory. On the other hand, the singular value decomposition of unitary matrices is not particularly helpful when dealing with unitary matrices, as for a complex-valued matrix $A$, the SVD requires the computation of the spectral properties of $AA^{\dagger}$ and $A^{\dagger}A$, which for any unitary matrix $A$ is always equal to the identity matrix. For these reasons, the density criterion discussed below suits better for the interlaced architectures here constructed.

To better understand the importance of dense matrices in our universal architecture, let us further inspect the factorization in~\eqref{eq_U}. By fixing $F=diag(e^{i \xi_{1}},\ldots, e^{i \xi_{N}})$ as a diagonal unitary matrix, for $\xi_{j}\in(0,2\pi]$ for $j=1,\ldots,N$, it is straightforward to notice that~\eqref{eq_U} reduces to a diagonal unitary matrix as well, which is far from representing a universal device. That is, diagonal $F$ matrices can only reconstruct diagonal unitary matrices. This behavior extends to any of the $N!^2$ different permutations allowed to the diagonal matrix $F$, as the factorized matrix $\mathcal{U}$ acquires the same structure as that of the permuted $F$ matrix. Now, let us consider the set of unitary matrices $\{\mathcal{V}_{n_{j}}\}_{j=1}^{k}$, where $\mathcal{V}_{n_{j}}\in U(n_{j})$ and $\sum_{j=1}^{k}n_{j}=N$, such that $F=diag(\mathcal{V}_{n_{1}},\ldots,\mathcal{V}_{n_{k}})$ is an $N$-dimensional block-diagonal unitary matrix. This particular selection for $F$ leads to a unitary operator with the same structure, which lacks the required universality property.  Although block-diagonal matrices $F$ lose their structure when randomly rearranged, anti-diagonal block matrices will always result in either block-diagonal or anti-block-diagonal matrices, neither of which are universal.

Thus, we have identified a particular class of \textit{bad-performing matrices} interlacing matrices $F$. Still, the latter is still useful to trace a suitable definition of density. To this end, let us first remark that any $N$-dimensional unitary matrix can be written as $\mathcal{V}=(\vec{v}_{1},\ldots,\vec{v}_{N})$, where $\vec{v}_{j}\in\mathbb{C}^{N}$ are complex-valued column vectors that form an orthogonal set through the Euclidean inner product in $\mathbb{C}^{N}$, i.e., $\vec{v}_{j_{1}}\cdot\vec{v}_{j_{2}}\equiv\vec{v}_{j_{1}}^{\dagger}\vec{v}_{j_{2}}=\delta_{j_{1},j_{2}}$. From the orthogonality condition, we can simply focus on the columns (or equivalently the rows) of $\mathcal{V}$, whereas the normalization imposes a constraint on elements across the columns (rows). Since the elements of $\mathcal{V}$ are complex numbers, we will work with the alternative matrix $\widetilde{\mathcal{V}}$ which is composed of the modulus of the elements of $\mathcal{V}$. Let us define $v_{p;q}$ as the $q$-th element of the $p$-th column vector $\vec{v}_{p}$, with $p,q=1,\ldots, N$, so that $\widetilde{\mathcal{V}}_{p;q}=\vert v_{p;q}\vert$. From the unitarity of $\mathcal{V}$, it follows that $\sum_{q=1}^{N}\vert v_{p;q}\vert^{2}=\sum_{p=1}^{N}\vert v_{p;q}\vert^{2}=1$ for all $p,q$, so that we can focus on the density of either the columns or rows of $\widetilde{\mathcal{V}}$. Without loss of generality, we work with the columns. Since we are interested in how the elements are sparse across each column, we compute the corresponding variance 
\begin{linenomath}
\begin{equation*}
S_{p}=\frac{1}{N}-\mu_{p}^{2}, \quad \mu_{p}=\frac{\sum_{q=1}^{N}\vert v_{p;q}\vert}{N}\geq \frac{1}{N} ,
\end{equation*}
\end{linenomath}
for each column $p$. Following the normalization condition, it is straightforward to prove that the variance is a bounded quantity in the interval $S_{p}=\left[0,\frac{N-1}{N^{2}}\right]$, where the lower bound corresponds to the case where all the elements of $\vec{\widetilde{v}}_{p}$ are equal, i.e., $\vec{\widetilde{u}}_{p}$ is maximally spread. The upper bound corresponds to the case where $\vec{\widetilde{v}}_{p}$ is one of the canonical unit vectors $(0,\ldots,1,\ldots,0)^{T}$. Thus, a given column $p$ of $\mathcal{V}$ is said to be denser if its corresponding variance $S_{p}$ approaches $(N-1)/N^2$. The more sparse the elements of the column $p$, the more $S_{p}$ approaches $0$. These are the key ideas we use henceforth to characterize density across the full matrix $\mathcal{V}$.

Be $\mathcal{S}=\{S_{p}\}_{p=1}^{N}$ the set of variances associated with each column of $\widetilde{\mathcal{V}}$. We define the mean $\widetilde{\mu}$ and standard deviation $\widetilde{\sigma}$ associated with the elements of $\mathcal{S}$, so that the density of a given unitary matrix $\mathcal{V}$ can be characterized by defining the point 
\begin{linenomath}
\begin{equation}
\label{crit-R}
\vec{R}:=(N\widetilde{\mu},N\widetilde{\sigma})) .
\end{equation}
\end{linenomath}
Since row permutation leaves $S_{p}$ invariant and column permutation only permutes the index $p$, the quantities $\widetilde{\mu}$ and $\widetilde{\sigma}$, and consequently the point $\vec{R}$, are permutation invariant. 

There are two note-worthy extremal cases, namely the maximally sparse and the diagonal cases (densest cases) unitary matrices. In the former case, $\mathcal{V}$ is composed of column vectors so that the variances vanish, $S_{p}=0$, for all $p=0,\ldots,N$. The DFT matrix of dimension $N$ is such an example. This leads to $\widetilde{\mu}=\sigma=0$, which we consider the ideal case. For the second case, the variances are maximal per each column, $S_{p}=(N-1)/N^{2}$, for all $p=1,\ldots,N$, and the statistical information of the matrix reduces to $\widetilde{\mu}=(N-1)/N^{2}$ and $\widetilde{\sigma}=0$. We thus have two comparison points, from which we find the bounded interval $N\widetilde{\mu}\in[0,(N-1)/N]$.

Additional reference points can be traced out if we take the block-diagonal matrices $F=diag(\mathcal{V}_{n_{1}},\ldots,\mathcal{V}_{n_{k}})$, with $\mathcal{V}_{n_{j}}\in U(n_{j})$ unitary and maximally sparse (DFT) matrices of dimension $n_{j}$ for $j=1,\ldots,k$, $1\leq k\leq N$, and $\sum_{j=1}^{k}n_{j}=N$, so that bad-performing matrices are generated (see discussion above). Particularly, let us consider the case $k=2$ so that $F=diag(\mathcal{V}_{n_{1}},\mathcal{V}_{n_{2}})$, with $n_{1}+n_{2}=N$. One can assign the indexes $n_{1}=\ell$ and $n_{2}=N-\ell$, with $\ell=1,\ldots,\lfloor N/2 \rfloor$ nonequivalent ways to define the two block-diagonal matrices $F$ that leads to the $k_2$ reference points
\begin{linenomath}
\begin{equation*}
\vec{R}_{k=2}=\left( \frac{2\ell (N-\ell)}{N^{2}},\frac{\sqrt{\ell(N-\ell)}(N-2\ell)}{N^{2}} \right) , \quad \ell=1,\ldots, \lfloor \frac{N}{2} \rfloor .
\end{equation*}
\end{linenomath}
Indeed, further reference points exist for $k\geq 3$. However, those points are farther from the ideal (maximally sparse) case $\vec{R}_{0}\equiv\Vec{R}_{k=1}$ than those marked with $k=2$. In this form, we can just focus on the area spanned between $k=1$ and $k=2$. Interestingly, for $k=2$ and $\ell=\lfloor N/2 \rfloor$, one obtains the reference points with smaller standard deviation, which are $\Vec{R}_{k=2,\ell=\frac{N}{2}}=\left( \frac{1}{2},0 \right)$ and $\Vec{R}_{k=2,\ell=\frac{N-1}{2}}=\left( \frac{1}{2}-\frac{1}{2N^{2}}, \frac{\sqrt{N^2-1}}{2N^2} \right)$ for even and odd $N$, respectively. Note that in the limit $N \rightarrow\infty$, the mean converges to the non-vanishing value $1/2$. In turn, the maximum standard deviation is determined by minimizing $N\widetilde{\sigma}$ in terms of $\ell$, from which one obtains the critical value $\ell_{c}=\frac{N}{2\sqrt{2}}(\sqrt{2}-1)$ and the maximum standard deviation $N\widetilde{\sigma}\vert_{\ell_{c}}=1/4$. That is, the standard deviation is bounded to the interval $N\widetilde{\sigma}\in [0,1/4]$, where the upper bound is independent of $N$ and is given by $max\left(N\widetilde{\sigma}\vert_{\lfloor \ell_{c} \rfloor}, N\widetilde{\sigma}\vert_{\lceil \ell_{c} \rceil} \right)$. The latter allows us forming a polygon with vertices at $\vec{R}_{k=1}=(0,0)$ and $\vec{R}_{k=2}$, the area of which is non-null and finite even for $N\rightarrow \infty$ (see Figure~\ref{fig3}). We focus on unitary matrices whose vector $\Vec{R}$ lies inside the latter polygon while avoiding the vertices, as the latter are the well-known bad-performing cases (with the exception of $\vec{R}_{0}$). We can go a step further and make a better prediction of unitary matrices by reducing the area of the polygon and imposing a threshold to $N\widetilde{\sigma}$ and $N\Vec{\mu}$. For the former, we already know that the standard deviation reaches its maximum value at $N\widetilde{\sigma}\vert_{k=2,\ell_{\sigma}}=1/4$ for all $N$. We thus implement the threshold at half of the maximum allowed standard deviation, i.e., $N\widetilde{\sigma}_{th}=1/8$. For even $N$, the reference points $\vec{R}_{k=2,\ell\geq\ell_{\sigma}}$ are above such a threshold for $\ell_{\sigma}=\frac{N}{2}-\lceil \frac{N}{4}\sqrt{2-\sqrt{3}} \rceil$. Likewise, we fix the threshold for the mean at $N\widetilde{\mu}_{th}=N\widetilde{\mu}\vert_{k=2,\ell=\ell_{\sigma}}=2\ell_{\sigma}(N-\ell_{\sigma})/N^{2}$. 

Therefore, any unitary matrix $F$ whose associated vector $\vec{R}$ lies inside the interception of the polygon spanned by the set of points $\{\vec{R}_{0}\}\cup\{\vec{R}_{k=2}\}_{\ell=1}^{\lfloor\frac{N}{2} \rfloor}$ and the thresholds $N\widetilde{\mu}_{th}$ and $N\widetilde{\sigma}_{th}$ has statistical properties in its matrix elements to be considered sparse enough to induce universality in our architecture. This provides a tool to quantify and classify \textit{a priori} the $F$ matrices. 


\subsection*{Random matrices and performance test}
\label{sec:measure-test}
To test this measure, we randomly generate families of $6\times 6$ unitary matrices and determine their location in the plane $(N\widetilde{\mu},N\widetilde{\sigma})$. The random matrices are generated by using the decomposition $F_{X}=e^{i X}$, with $X^{\dagger}=X$ a Hermitian matrix in $\mathbb{C}^{N\times N}$ to be defined. In order to preserve control over the density of the randomly generated matrices, we introduce the four non-symmetric matrices $X_A=(\vec{\chi}_{1;A},\vec{0},\vec{0},\vec{0},\vec{0},\vec{0})$, $X_B=(\vec{\chi}_{1;B},\vec{\chi}_{2;B},\vec{0},\vec{0},\vec{0},\vec{0})$, $X_C=(\vec{\chi}_{1;C},\vec{\chi}_{2;c},\vec{\chi}_{3;C},\vec{0},\vec{0},\vec{0})$, and $D=(\vec{\chi}_{1;D},\vec{\chi}_{2;D},\vec{\chi}_{3,D},\vec{\chi}_{4;D},\vec{0},\vec{0})$, with $\vec{0}$ the null-vector in $\mathbb{C}^{N}$. The column vectors $\vec{\chi}_{j;\wp}$ are composed of zeros in the first $j$ inputs and random numbers elsewhere, with $\wp\in\{A,B,C,D\}$. We thus consider the Hermitian construction as $X=X_{\wp}+X_{\wp}^{\dagger}$, with $\wp\in\{A,B,C,D\}$. Note that in each case, the number of random parameters increases in each case as additional columns are included, which implies that more parameters are available to build up the corresponding unitary matrices. We thus expect denser matrices constructed from $D$ than $A$. 

In this form, we establish a controlled testing scenario for our density criterion and the subsequent universality of the matrix under consideration. These results are depicted in Figure~\ref{fig3}(a), where the reference points $\Vec{R}_{k}$ and the points $\vec{R}$ associated with the sets of random unitary matrices $F_{X_{\wp}}$ for $\wp\in\{A,B,C,D\}$, so that we can determine the unitary matrices whose corresponding points $\vec{R}$ lie inside the universality area. On the one hand, for the random unitary matrices $F_{A}$, only one of the points, $A_{5}$, is expected to reveal universality properties. On the other hand, we expect more matrices $F_{D}$ with such a universality, such as $D_{1}$, $D_{5}$, and $D_{12}$, to name a few; whereas the matrices associated with $D_{3}$, $D_{10}$, and $D_{19}$ are expected to lack the desired property. This criterion is tested by using the latter unitary matrices, $F_{\ell}$, and testing their performance for randomly generated unitary target matrices. Here, fifty target unitary matrices are considered for each testing unitary matrix so that the relative error of the so-reconstructed targets can be analyzed for a broad number of cases (see Figure~\ref{fig3}(b)). Such a performance test reveals a good match with our universality criterion. For instance, the error associated with $F_{A_{2}}$ and $F_{A_{6}}$ is relatively high for each of the target matrices tested; in turn, $F_{A_5}$ performed well in each scenario, and $F_{A_{13}}$ and $F_{A_{20}}$ have revealed a handful of cases with a higher error. These two matrices lie far outside the universality area shown in Figure~\ref{fig3}(a), and thus a bad performance was expected. The performance for the unitary matrices $F_{D_{j}}$ reveals a better fit across all the cases, where only $F_{D_{3}}$, $F_{D_{10}}$ and $F_{D_{19}}$ contain a few cases in which the target is reconstructed with a moderate error around $10^{-5}$. 

As the number of input channels in an architecture increases, the time required for performance testing also increases. Therefore, the density criterion allows us to classify and select in advance the required $F$ matrices that are expected to lead to universal behavior. Another advantage is that the proposed universality area spanned in the $\Vec{R}$-space is finite and non-null for $N\rightarrow\infty$, making it a suitable measure for architectures with an arbitrary number of ports. Thus, in Figure~\ref{fig3}(c), we depict a more convenient representation of our results in which we represent both $N\widetilde{\mu}$ and $N\widetilde{\sigma}$ separately, where we determine universality if both quantities lie below their respective thresholds. This permits a better and more organized reading for each matrix $F$ under consideration. Remark that, although our density criterion determines the boundaries where universality is likely, it may exclude some well-performing cases, i.e., this is a sufficient but not necessary condition to identify universality. 



\subsection*{Photonic Realizations}
So far, the universality of the proposed architecture has been successfully determined using Haars random $F$ matrices possessing the required density criterion. We now proceed to analyze potential candidates for photonic realizations of such F matrices, with a primary focus on photonic lattices and meshes of directional couplers.

\subsubsection*{Waveguide Lattices}
Waveguide arrays can be modeled with high precision using coupled-mode theory~\cite{Huang94}. The latter takes into account the coupling of evanescent waves from one waveguide interacting with its nearest-neighbor while neglecting farther neighbors due to their weak coupling. In this form, the effective Hamiltonian describing an array of $N$ waveguides is characterized by a tridiagonal and symmetric matrix Hamiltonian $\mathbb{H}$ of dimension $N$. The wave evolution through the lattice is ruled by the dynamical law $i\frac{d}{dz}\vec{u}(z)=\mathbb{H}\cdot\Vec{u}(z)$, where $\Vec{u}(z)\in\mathbb{C}^{N}$ is the electric field at each waveguide at the propagation distance $z$. Since $\mathbb{H}\neq\mathbb{H}(z)$, the wave evolution is determined through the unitary evolution operator $\mathbb{F}(z)=e^{-iz\mathbb{H}}$ as $\vec{u}(z)=\mathbb{F}(z)\Vec{u}(z=0)$. 

\begin{figure*}
    \centering
    \includegraphics[width=0.9\textwidth]{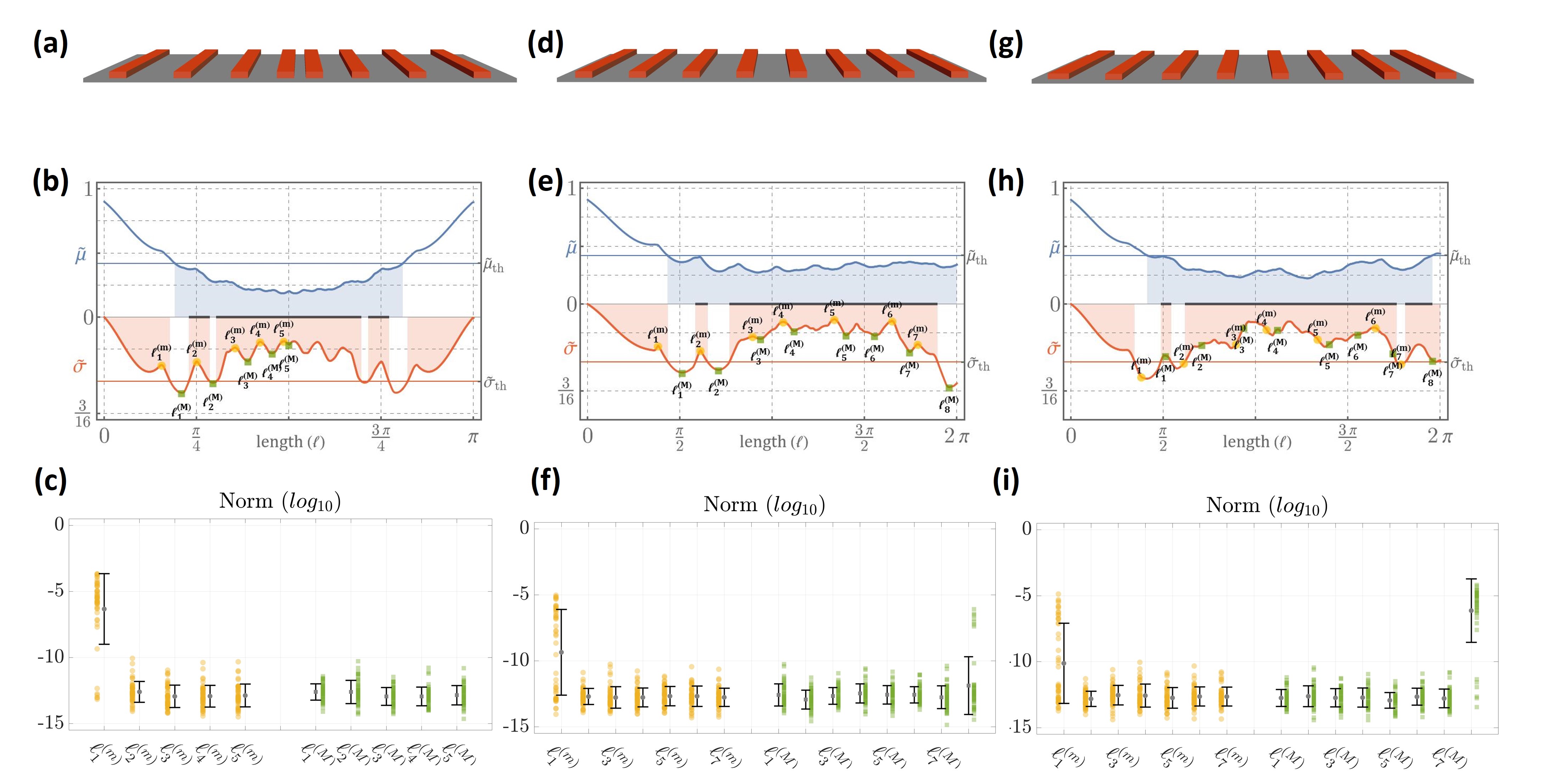}
    \caption{\textbf{Photonic lattice realization and their universality.} Sketch for the waveguide array associated with the $J_{x}$ lattice (a), homogeneous lattice (b), and homogenous lattice with disorder effects (c). Density criterion as a function of the lattice length $\ell$ (d-f) and the corresponding numerical performance test at the reference lengths $\ell^{(m)}_{j}$ and $\ell^{(M)}_{j}$ (g-i) for the $J_x$ lattice (left column), homogeneous lattice (central column), and disordered lattice (right column).}
    \label{fig4}
\end{figure*}

We can thus implement waveguide arrays in the universal architecture, provided they fulfill the desired universality. To this end, we can test the behavior of a given lattice evolution operator for specific lengths using the density criteria of Section~\ref{sec:density}. Particularly, we consider the photonic $J_{x}$ lattice~\cite{Wei16}, the homogeneous lattice~\cite{Chr03}, and the disordered homogeneous lattice~\cite{Miri21} as the physical waveguide arrays under consideration described by the respective Hamiltonians $\mathbb{H}^{(J_{x})}$, $\mathbb{H}^{(h)}$, and $\mathbb{H}^{(h,d)}$. The matrix elements of the latter are explicitly given by
\begin{linenomath}
\begin{equation}
\label{Hs}
\begin{aligned}
& \mathbb{H}^{(J_{x})}_{p,q}:=\kappa(p)\delta_{p+1,q}+\kappa(p-1)\delta_{p-1,q} , \\
& \mathbb{H}^{(h)}_{p,q}:=\kappa_{0}\delta_{p+1,q}+\kappa_{0}\delta_{p-1,q} , \\
& \mathbb{H}^{(h,d)}_{p,q}:=(\kappa_{0}+\Delta\kappa_{p})\delta_{p+1,q}+(\kappa_{0}+\Delta\kappa_{p-1})\delta_{p-1,q} ,
\end{aligned}
\end{equation}
\end{linenomath}
with $p,q\in\{1,\ldots,N\}$. Here, $\kappa(p)=\frac{\kappa_{0}}{2}\sqrt{(N-p)p}$ stands for the coupling parameter between nearest waveguide neighbors in the $J_x$ lattice, whereas $\Delta\kappa_p\in N(\mu,\sigma)$ are random numbers taken from the \textit{normal distribution} $N(\mu,\sigma)$ characterizing the disorder effects. The coupled waveguide implementation for each Hamiltonian is depicted in Figure~\ref{fig4}.

\begin{table}[!ht]
\centering
\caption{Lattice lengths at the local minima $\ell_{j}^{(m)}$ and local maxima $\ell_{j}^{(M)}$ of the density criterion $N\widetilde{\sigma}$ for the $J_{x}$ and homogeneous lattice.}
\begin{tabular}{c|c|c|c|c|c}
       $z$  & $J_{x}$ & Hom. & $z$  & $J_{x}$ & Hom. \\ \hline
       $\ell_{1}^{(m)}$  & 0.48781 & 1.1968 & $\ell_{1}^{(M)}$  & 0.6579    & 1.6099 \\
       $\ell_{2}^{(m)}$  & 0.78735 & 1.9211 & $\ell_{2}^{(M)}$  & 0.9256    & 2.2285 \\
       $\ell_{3}^{(m)}$  & 1.11393 & 2.8088 & $\ell_{3}^{(M)}$  & 1.2227    & 2.9441 \\
       $\ell_{4}^{(m)}$  & 1.32647 & 3.3290 & $\ell_{4}^{(M)}$  & 1.4293    & 3.5156 \\
       $\ell_{5}^{(m)}$  & 1.52812 & 4.1988 & $\ell_{5}^{(M)}$  & $\pi/2$   & 4.3984 \\
       $\ell_{6}^{(m)}$  &         & 5.1835 & $\ell_{6}^{(M)}$  &           & 4.8832 \\
       $\ell_{7}^{(m)}$  &         & 5.6182 & $\ell_{7}^{(M)}$  &           & 5.4811 \\
                         &         &        & $\ell_{8}^{(M)}$  &           & 6.1544
\end{tabular}
\label{tab:lengths}
\end{table}

The corresponding unitary evolution operators are simply given by $\mathbb{F}^{(J_{x})}(z)=e^{iz\mathbb{H}^{(J_{x})}}$ and $\mathbb{F}^{(h)}(z)=e^{iz\mathbb{H}^{(h)}}$. Although both are functions of the lattice length $z$, the $J_x$ lattice (Figure~\ref{fig4}(a)) has equidistant eigenvalues that lead to a periodic unitary evolution operator $\mathbb{F}^{(J_{x})}(z)$ in $z$, so that we can simply focus on the interval $z\in [0,2\pi)$. We first estimate the lengths that induce universality in our architecture, which is depicted in Figure~\ref{fig4}(b). In the latter, we mark the particular lengths $z_{j}^{(m)}$ and $z_{j}^{(M)}$ that denote the local minima and maxima of the standard deviation $N\widetilde{\sigma}$, the exact values of which have been determined numerically and presented in Table~\ref{tab:lengths}. In turn, the black-thick line in Figure~\ref{fig4}(b) denotes the lengths where both $N\widetilde{\mu}$ and $N\widetilde{\sigma}$ are below the universality threshold; i.e., the lengths where universality is expected. Without any prior performance test, one can see that the lengths $z^{(m)}_{1}$, $z^{(M)}_{1}$, and $z^{(M)}_{2}$ may fail in obtaining the desired universality. Recall that our estimation criterion is only a sufficient condition and may rule out positive cases. Nevertheless, all the other marked points can be considered candidates for the matrix $F$ in our architecture, as no false positive cases will be included. This is indeed verified in the performance test portrayed in Figure~\ref{fig4}(c), where, for each point, we have used fifty randomly generated unitary matrices as targets. The latter confirms our predictions, where the only bad-performing length is found at $z^{(M)}_{1}$, reinforcing the fact that only two positive cases were discarded, but no false positives were included. In this form, we can confidently conclude that a universal architecture can be built using $J_{x}$ lattices with lengths as small as $z=\pi/4$ for $N=10$.

We alternatively consider the homogeneous lattice, which contains waveguide arrays homogeneously distributed (Figure~\ref{fig4}(d)). The eigenvalues accordingly distributed as $\lambda_{n}^{(h)}=2\kappa_{0}\cos(\frac{n\pi}{N+1})$, and no periodic behavior is expected. We thus focus on the interval $\ell\in[0,4\pi]$ for this particular lattice. The density criterion shown in Figure~\ref{fig4}(e) reveals that lattice lengths in the interval $z/\kappa_{0}\in(2.4098,5.9595)$ are suitable for our universal architecture. Particularly, note that the interval $z/\kappa_{0}\in[\ell_{4}^{(m)},\ell_{6}^{(m)}]$ contains lengths so that $N\widetilde{\mu}$ and $N\widetilde{\sigma}$ remain mostly constant with minor variations. Thus, the performance test in this interval is expected to perform well. In the universality region, there is a local minimum $\ell_{2}^{(m)}$ that is isolated and associated with a shorter lattice length. This reference point may be useful for reducing the size of the universal structure. Figure~\ref{fig4}(f) displays the corresponding performance test, which supports our previous statements. As expected, the performance for lattices with length $\ell_{1}^{(m)}$ and $\ell_{8}^{(M)}$ is particularly poor. However, the length $\ell_{1}^{(M)}$ has a generally good performance, with only two test targets displaying slightly higher errors than the other well-performing cases.

We additionally take into account the effects of disorder on the homogeneous lattice, which may be caused by impurities or imperfections during the manufacturing process, resulting in waveguides not being in their ideal positions or displaying deviations in their sizes (see Figure~\ref{fig4}(g)). The defects here considered are such that the nearest-neighbor interactions deviate from the ideal homogeneous lattice by a factor of twenty percent; i.e., the disorder couplings in~\eqref{Hs} take values from the normal distribution as $\Delta\kappa_p\in N(\mu=0,\sigma=0.2\kappa_{0})$. Although the lattice structure is modified in the latter disorder, the estimation of density does not differ significantly from the ideal case, as shown in Figures~\ref{fig4}(h)-(i). 

\subsubsection*{Directional Coupler Mesh}

\begin{figure*}
    \centering
    \includegraphics[width=0.8\textwidth]{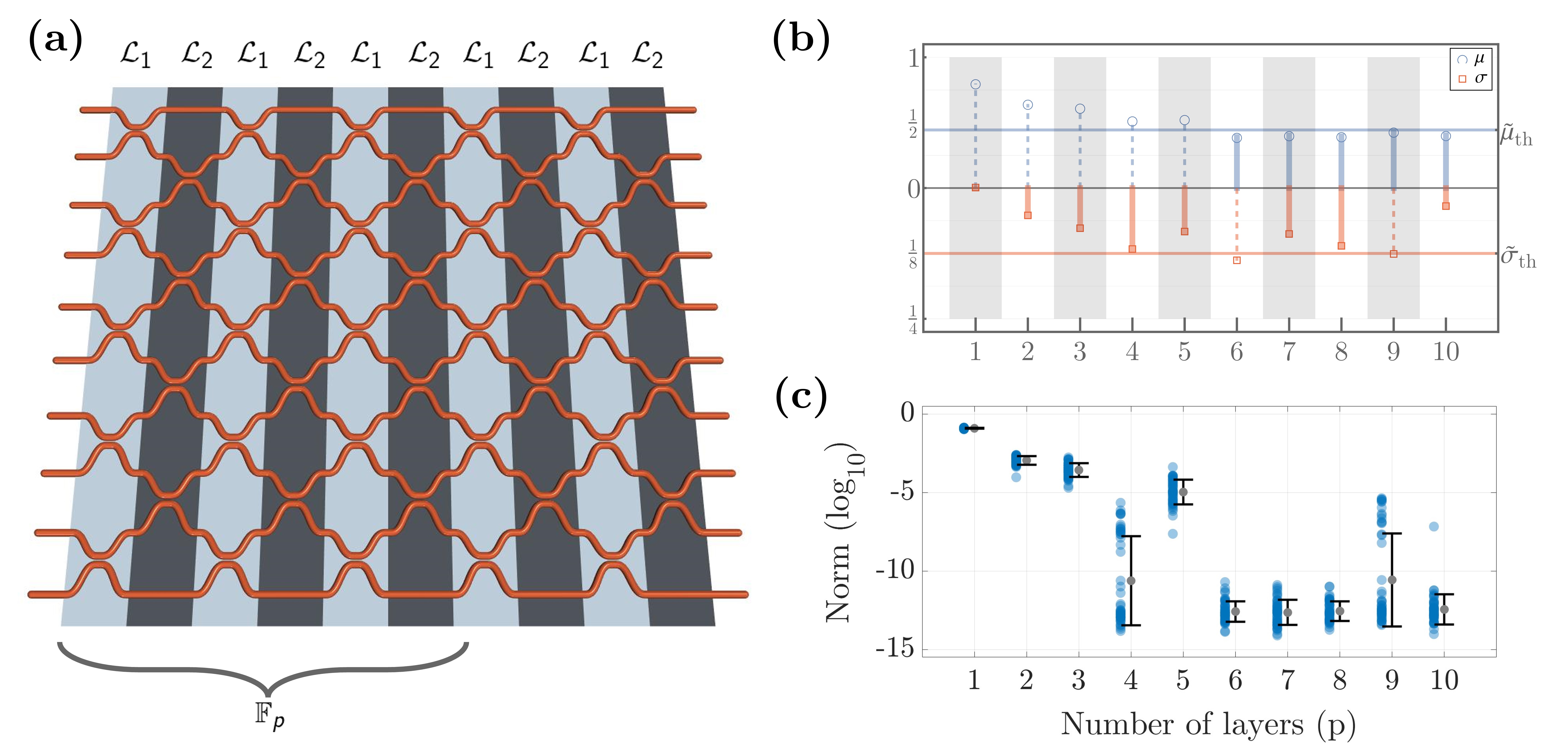}
    \caption{\textbf{Geometric array using power dividers.} (a) Power divider array composed of $p$ layers as defined in~\eqref{PD}. Light-shaded and dark-shaded layers denote $\mathcal{L}_{1}=\mathbb{I}_{5}\otimes\mathcal{T}_{0}$ and $\mathcal{L}_{2}=\mathbb{I}_{1}\oplus(\mathbb{I}_{4}\otimes\mathcal{T}_{0})\oplus\mathbb{I}_{1}$, respectively. Density criterion (b) and error norm (log$_{10} L$) (c) of the mesh architecture in (a) as a function of the number of layers $p$.}
    \label{fig5}
\end{figure*}

Alternatively, the matrix $F$ can be optically realized through proper mesh arrays of directional couplers. Particularly, we consider a construction based on two-port passive elements, which act as a power divider (50:50 MZI) equivalent, up to a global phase, to the unitary matrix $U(2)\ni\mathcal{T}_{0}=\frac{1}{\sqrt{2}}\left(\sigma_{0}+i\sigma_{1}\right)$, with $\sigma_{j}$ the conventional Pauli matrices for $j\in\{1,2,3\}$ and $\sigma_{0}$ the $2\times 2$ identity matrix. The latter can be used as a building block to construct other $U(2)$ matrices~\cite{deGuise18} as well as higher dimensional unitary matrices $U(N)$ through appropriate Kronecker products~\cite{Reck94}. In this section, we consider the symmetric $10$-port array portrayed in Figure~\ref{fig5}(a), composed of power dividers $\mathcal{T}_{0}$ interconnected through different layers $\mathcal{L}_{1}$ and $\mathcal{L}_{2}$. Here, each layer is described by the $U(10)$ matrices $\mathcal{L}_{1}=\mathbb{I}_{5}\otimes\mathcal{T}_{0}$ and $\mathcal{L}_{2}=\mathbb{I}_{1}\oplus(\mathbb{I}_{4}\otimes\mathcal{T}_{0})\oplus\mathbb{I}_{1}$, with $\otimes$ and $\oplus$ the Kronecker product and Kronecker sum\footnote{Such operations are also known as direct product and direct sum.}, respectively, and $\mathbb{I}_{n}$ the $n\times n$ identity matrix. The $p$-layered unitary matrix describing the power divider array is thus given by 
\begin{linenomath}
\begin{equation}
\label{PD}
\mathbb{F}_{p}=\underbrace{\mathcal{L}_{\widetilde{p}}\ldots\mathcal{L}_{1}\mathcal{L}_{2}\mathcal{L}_{1}}_{p\textnormal{-times}}, \quad 
\widetilde{p}=
\begin{cases}
1, & p\in \{1,3,5,7,9 \} \\
2, & p\in \{2,4,6,8,10\}
\end{cases}
,
\end{equation}
\end{linenomath}
where we have truncated the maximum number of layers to ten. 

It is not mandatory to truncate these layers, and the procedure can involve additional layers if needed. However, for practical physical implementations and limitations, we aim for compact devices and need to limit the number of layers as much as possible. To this end, we estimate the number of layers $p$ for which universality is achieved. The preliminary estimation displayed in Figure~\ref{fig5}(b) ensures the goodness of $F$ for $p=7,8,10$, being $p=10$ the case in which both $N\widetilde{\mu}$ and $N\widetilde{\sigma}$ minimize. Based on the performance test shown in Figure~\ref{fig5}(c), it was found that the $p=7,8,10$ layers are capable of achieving a $F$ capable of achieving the required universality in the architecture~\eqref{eq_U}. Notably, optimization results indicate that our device could perform well with only $p=6$ layers, which was originally deemed inadequate from the density criterion. That is, we obtain a false negative for $F$. Still, no false positives were detected during the analysis, which is strictly necessary to avoid faulty designs of the final architecture.


\section*{Conclusions}

We have introduced the design for a lossless universal photonic architecture based on a layered scheme of interlaced active phase shifter layers and passive random matrices. Numerical results obtained from the LMA optimization revealed that generating Haar random matrices $F$ leads, in a vast majority of cases, to the desired universal architecture. It is observed that well-behaved matrices $F$ show a phase transition on the error norm $L$ of the reconstructed target at $M=N+1$, with $M$ the total number of phase shifter layers. In such a layer number, the error drops significantly to numerical noise values. While this is not proof that the factorization is exact, the error involved in the reconstruction process lies in the numerical error regime, and it is thus low enough to ensure that any unitary matrix is reconstructed with the desired accuracy.

Despite the accuracy of the LMA optimization, the computational time required for testing the universality of the random matrices $F$ scales with the total number of ports, which becomes impractical for particularly large architectures. Numerical evidence shows that denser matrices perform better than sparse ones, usually involving relatively large errors. Therefore, a density criterion has been devised and introduced to classify the candidates for the matrix $F$ used in the architecture. This criterion is built on preliminary knowledge of bad-performing matrices, such as diagonal and block diagonal matrices, which are analytically known to fail but serve as reference points to look for good-performing matrices, such as the DFT case. In this form, instead of performing a long optimization routine on the candidate for $F$, we simply analyze the standard deviation of the modulus of its columns or rows, which provides information about its density. This allows defining a mapping $\vec{R}:U(N)\rightarrow\mathbb{R}^{2}$, which renders a vector that estimates whether $F$ is suitable for the architecture. We thus possess a tool to preselect matrices $F$ beforehand, making the design process more practical than generating and testing several random matrices.

Our tests using randomly generated unitary matrices showed that matrices within the threshold marked by the density criterion led to the required universality. Thus, universality is not limited to a specific realization of $F$; as shown in the results, infinitely many unitary matrices can meet our requirements. This paves the way for more efficient construction and optimization of compact devices that are simultaneously resilient to random defects. Particularly, the photonic Jx lattice was found suitable for this task at lengths different than the previously reported critical value $\ell=\pi/2$~\cite{markowitz2023universal}. This defines intervals in the lattice length for which the architecture is universal, leading to more flexibility in the manufacturing process so that one can allow for deviations in lattice length. This fact is further supported in the context of homogenous lattices, which are also suitable for our architecture and robust against disorder effects due to waveguide impurities or mismatching sizes. The latter was tested by introducing deviations of up to $20\%$ into the homogeneous lattice, from which the density estimation showed no significant difference in the universality performance for the lattice lengths considered. Further constructions for the $F$ matrices are indeed allowed, and an alternative construction based on a layered array of power dividers was shown to be efficient for our purposes, the analysis of which allowed us to determine the optimal number of passive elements required for the architecture. 

The presented results, in particular the density criterion, are not limited to the waveguide and direction coupler realization discussed in the manuscript. Indeed, any optical element described by a unitary matrix, such as multimode interference couplers, can be further tested by the density criterion and classified before being used in the layered architecture~\eqref{eq_U}. The latter helps facilitate the design of the passive layer to reduce the overall architecture size, account for potential manufacturing errors, and diminish the device footprint. This is particularly handy when deploying more complex optical circuits.



\subsection*{Acknowledgments}
This project is supported by the U.S. Air Force Office of Scientific Research (AFOSR) Young Investigator Program (YIP) Award\# FA9550-22-1-0189 and the City University of New York (CUNY) Junior Faculty Research Award in Science and Engineering (JFRASE) funded by the Alfred P. Sloan Foundation.

\subsection*{Author Contributions}
All authors have accepted responsibility for the entire content of this manuscript and approved its submission.

\subsection*{Conflict of Interest}
Authors state no conflict of interest.

\subsection*{Data Availability}
The datasets generated during and/or analyzed during the current study are available from the corresponding author on reasonable request.

\subsection*{Disclosures} 
Patent Pending.

\bibliography{biblio}

\end{document}